\documentclass[sigconf, screen, nonacm]{acmart}

\usepackage{tikz}
\usetikzlibrary{patterns,decorations.pathreplacing, fadings}
\definecolor{lightblack}{RGB}{33, 33, 33}
\definecolor{lightgray}{RGB}{240, 240, 240}
\definecolor{lightorange}{RGB}{255, 245, 242}
\definecolor{lightblue}{RGB}{194, 231, 255}

\usepackage{hyperref}
\definecolor{linkcolor}{RGB}{18, 77, 150}
\bibpunct{\color{linkcolor}[}{\color{linkcolor}]}{,}{n}{}{;}
\newcommand{\reffig}[1]{\hyperref[#1]{\textcolor{linkcolor}{Figure \ref*{#1}}}}
\newcommand{\reftab}[1]{\hyperref[#1]{\textcolor{linkcolor}{Table \ref*{#1}}}}
\newcommand{\refsec}[1]{\hyperref[#1]{\textcolor{linkcolor}{Section \ref*{#1}}}}

\begin{document}

\title[Real-Time Forecasting of Keyboard and Mouse Actions]{A Click Ahead: Real-Time Forecasting of Keyboard and Mouse Actions using RNNs and Computer Vision}

\author{Fabio Matti}
\email{fabio.matti@epfl.ch}
\affiliation{%
  \institution{Ecole poyltechnique fédérale de Lausanne (EPFL)}
  \city{Lausanne}
  \country{Switzerland}
}

\author{Pierre Dillenbourg}
\email{pierre.dillenbourg@epfl.ch.}
\affiliation{%
  \institution{Ecole poyltechnique fédérale de Lausanne (EPFL)}
  \city{Lausanne}
  \country{Switzerland}
}

\author{Ludovico Novelli}
\email{lnovelli@logitech.com}
\affiliation{%
  \institution{Logitech International S.A.}
  \city{Lausanne}
  \country{Switzerland}
}

\begin{abstract}
Computer input is more complex than a sequence of single mouse clicks and keyboard presses. We introduce a novel method to identify and represent the user interactions and build a system which predicts -- in real-time -- the action a user is most likely going to take next. For this, a recurrent neural network (RNN) is trained on a person's usage of the computer. We demonstrate that it is enough to train the RNN on a user's activity over approximately a week to achieve an accuracy of 34.63 \% when predicting the next action from a set of almost 500 possible actions. Specific examples for how these predictions may be leveraged to build tools for improving and speeding up workflows of computer users are discussed.
\end{abstract}

\maketitle

\section{Introduction}

When working on the computer users tend to repeat the same input patterns over and over again. This motivates us to explore ways in which these repetitions can be learned and used to predict what the user is going to do next. These predictions can subsequently be leveraged to assist the user in multiple ways which we will discuss at the end of this paper.

There are many examples of systems which model the users' inputs while performing a specific subset of tasks: \citet{MouseKeyboardGazeIntentTextFormatting} predict the users' next action while formatting text in a simplified text editor; \citet{MouseCursorMovementAttentionDisplayLSTMGRU} predict user attention in search engine result pages; \citet{ItemPurchaseIntentLSTM} learn intent on online marketplaces; and \citet{GazeBasedIntentPrediction} focuses on predicting button presses while solving a puzzle. We extend the scope by taking -- what to our knowledge has not been attempted yet -- a task- and app-agnostic approach to input modeling.

Contrary to systems which use simulated environments with which the user interacts \cite{MouseCursorMovementAttentionDisplayLSTMGRU, ItemPurchaseIntentLSTM}, we need to face the challenge of determining what effect a mouse event has on the state of an application or website. It is of little use to save the absolute position of the cursor on the screen, because the position of the button on the screen often varies depending on the positioning of the window, the size of the window, and the screen's DPI. Hence, we resort to a computer vision-based approach to identify which interactive area on the current screen was clicked (button, link, text field...). We do this by assembling a database containing image patches of all the interactive areas (often simply referred to as buttons) clicked by the user, which can be searched through and extended on the fly. Collecting these image patches also allows us to reliably reenact actions involving clicks on interactive areas. 

Some examples in literature also account for gaze dynamics in their predictions \cite{MouseKeyboardGazeIntentTextFormatting, GazeBasedIntentPrediction}. Due to privacy concerns, computational constraints, hardware requirements, robustness, and simplicity, we make no use of this information. 

The go-to models for time series forecasting are recurrent neural networks (RNN). Long short-term memories (LSTM) \cite{LSTM}
and gated recurrent units (GRU) \cite{GRU} appear particularly often in the context of predicting user intent \cite{MouseCursorMovementAttentionDisplayLSTMGRU, ItemPurchaseIntentLSTM}. Isolated appearances of hidden Markov models (HMM) are also seen in literature \cite{LearningAndAnticipatingMouseInteractionsHMM}.

Our main contribution is the general representation of computer inputs which we show to be efficiently learnable by RNNs. 

    \section{User input detection and representation}

To represent user input such that it is independent of which application the action is taken in and also sufficiently describes the action such that it can be automatically reenacted at a later time, we hereafter define the concept of a \textit{user action}. It is important to note that a good representation uses as few features as possible to describe the input, keeps the space of possible actions small, and is consistent between different instances of a user performing the same action.

\begin{figure}[ht]
    \centering
    \scalebox{0.23}{\begin{tikzpicture}
    \draw[->, ultra thick] (0, 0) to (35, 0) node[right, font={\Huge\bfseries\sffamily}] {time};

    \draw[ultra thick, dashed] (0, 4.5) to (35, 4.5);

    \node[rotate=90, anchor=center, font={\Huge\bfseries\sffamily}] at (0.25, 8) {USER ACTIONS};
    \node[rotate=90, anchor=center, font={\Huge\bfseries\sffamily}] at (0.25, 2.25) {INPUTS};

    \draw[->, ultra thick] (2, 0.25) to (2, 5);
    \fill[lightblack] (1.5, 5.25) rectangle (2.5, 11) node[midway, color=white, rotate=90, font={\Huge\bfseries\sffamily}] {C};
    
    \fill[lightblack] (2, 0.25) rectangle (3, 1.25) node[midway, color=white, font={\Huge\bfseries\sffamily}] {C};

    \draw[->, ultra thick] (3.2, 0.25) to (3.2, 5);
    \fill[lightblack] (2.7, 5.25) rectangle (3.7, 11) node[midway, color=white, rotate=90, font={\Huge\bfseries\sffamily}] {H};
    
    \fill[lightblack] (3.2, 0.25) rectangle (4.7, 1.25) node[midway, color=white, font={\Huge\bfseries\sffamily}] {H};

    \draw[->, ultra thick] (5, 0.25) to (5, 5);
    \fill[lightblack] (4.5, 5.25) rectangle (5.5, 11) node[midway, color=white, rotate=90, font={\Huge\bfseries\sffamily}] {I};
    
    \fill[lightblack] (5, 0.25) rectangle (5.6, 1.25) node[midway, color=white, font={\Huge\bfseries\sffamily}] {I};

    \draw[->, ultra thick] (8, 0.25) to (8, 5);
    \fill[lightblack] (7.5, 5.25) rectangle (8.5, 11) node[midway, color=white, rotate=90, font={\Huge\bfseries\sffamily}] {CTRL + C};
    \draw[->, ultra thick] (13, 0.25) to (13, 5);
    \fill[lightblack] (12.5, 5.25) rectangle (13.5, 11) node[midway, color=white, rotate=90, font={\Huge\bfseries\sffamily}] {CTRL + V};
    
    \fill[lightblack] (6, 0.25) rectangle (16, 1.25) node[midway, color=white, font={\Huge\bfseries\sffamily}] {CTRL};
    \fill[lightblack] (8, 1.5) rectangle (11.5, 2.5) node[midway, color=white, font={\Huge\bfseries\sffamily}] {C};
    \fill[lightblack] (13, 1.5) rectangle (15, 2.5) node[midway, color=white, font={\Huge\bfseries\sffamily}] {V};

    \draw[->, ultra thick] (17, 0.25) to (17, 5);
    \fill[lightblack] (16.5, 5.25) rectangle (17.5, 11) node[midway, color=white, rotate=90, font={\Huge\bfseries\sffamily}] {SPACE};
    
    \fill[lightblack] (17, 0.25) rectangle (20, 1.25) node[midway, color=white, font={\Huge\bfseries\sffamily}] {SPACE};

    
    \draw[->, ultra thick] (29, 0.25) to (29, 5);
    \fill[lightblack] (28.5, 5.25) rectangle (29.5, 11) node[midway, color=white, rotate=90, font={\Huge\bfseries\sffamily}] {CTRL + ALT + DEL};
    
    \fill[lightblack] (21, 0.25) rectangle (33, 1.25) node[midway, color=white, font={\Huge\bfseries\sffamily}] {CTRL};
    \fill[lightblack] (25, 1.5) rectangle (34, 2.5) node[midway, color=white, font={\Huge\bfseries\sffamily}] {ALT};
    \fill[lightblack] (29, 2.75) rectangle (32, 3.75) node[midway, color=white, font={\Huge\bfseries\sffamily}] {DEL};
\end{tikzpicture}}
    \caption{Conversion of raw key and modifier inputs to user actions. The black bars show the time span during which the corresponding key was held.}
    \label{fig:userinput}
\end{figure}
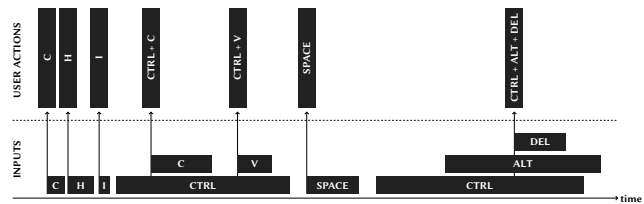

Simple keystrokes (such as characters, numbers, and navigation keys) can be trivially identified as their own user actions. However, modifier keys (such as CTRL, SHIFT, and ALT) acquire meaning only when used in combination with other keys. Thus, it is of no importance when a modifier key is pressed and when it is released, but only which modifier keys are active during a simple keystroke. Every encountered key-modifier combination (e.g. CTRL $+$ C,  CTRL $+$ V, and CTRL $+$ ALT $+$ DEL) is a distinct user action. \reffig{fig:userinput} illustrates how sequences of keyboard inputs are converted to user actions.

\begin{figure}[ht]
    \centering
    \scalebox{0.3}{\begin{tikzpicture}

    \begin{scope}[shift={(-13, 0)}]
 
        \fill[black!7!white] (-5, 1.5) rectangle (5, -3);
        \fill[black!1!white] (5, -3.5) to (-5, -3.5) to (-5, 0.9) to (0.5, 0.9) arc (90:0:0.3) to (0.8,-0.55) arc (180:270:0.3) to (5, -0.85) to cycle;
        \fill[black!6!white] (5, -1.15) to (1.4, -1.15) arc (90:270:0.9) to (5, -2.95) to cycle;
        \fill[black!15!lightblue] (-5, 1.5) rectangle (5, 3.0);
    
        \draw[black!80!white, line width=2] (-0.2, -0.2) to (0.2, 0.2);
        \draw[black!80!white, line width=2] (-0.2, 0.2) to (0.2, -0.2);
    
        \draw[black!80!white, line width=2] (-4.35, -2.0) to (-3.7, -2.0);
        \draw[black!80!white, line width=2] (-4.0, -1.7) to (-3.7, -2.0) to (-4.0, -2.3);
    
        \draw[black!80!white, line width=2] (-2.33, -2.0) circle (0.32);
        \fill[black!1!white] (-2.1, -1.95) rectangle (-1.8, -2.1);
        \fill[black!80!white] (-2.22, -1.95) to (-1.98, -1.95) to (-1.98, -1.7) to cycle;
    
        \draw[black!80!white, line width=2] (-0.88, -2.35) to (-0.88, -2.0) to (-0.95, -2.0) to (-0.65, -1.7) to (-0.35, -2.0) to (-0.43, -2.0) to (-0.43, -2.35) to cycle;
    
        \draw[black!80!white, line width=2.5] (1.65, -0.35) to (1.65, 0.35);
        \draw[black!80!white, line width=2.5] (1.3, 0.0) to (2.0, 0.0);
    
        \node[font={\Huge\bfseries\sffamily}, black!80!white] at (-3.37, 0.0) {New Tab};
    
        \node[font={\Huge\bfseries\sffamily}, black!50!white] at (2.5, -2.1) {Search};
    
        \begin{scope}[shift={(-14.9, 1.85)}]
            \draw[black, fill=white, line width=2.5pt] (15, -2) to (15, -3) to (15.25, -2.9) to (15.4, -3.25) to (15.6, -3.15) to (15.45, -2.8) to (15.7, -2.7) to cycle;
        \end{scope} 

        \draw[black] (-5, 3) rectangle (5, -3.5);

    \end{scope}

    \draw[->, line width=4] (-7, 0) to (-6, 0);

    \begin{scope}
    
        \fill[black!7!white] (-5, 1.5) rectangle (5, -3);
        \fill[black!1!white] (5, -3.5) to (-5, -3.5) to (-5, 0.9) to (0.5, 0.9) arc (90:0:0.3) to (0.8,-0.55) arc (180:270:0.3) to (5, -0.85) to cycle;
        \fill[black!6!white] (5, -1.15) to (1.4, -1.15) arc (90:270:0.9) to (5, -2.95) to cycle;
        \fill[black!15!lightblue] (-5, 1.5) rectangle (5, 3.0);
    
        \draw[black!80!white, line width=2] (-0.2, -0.2) to (0.2, 0.2);
        \draw[black!80!white, line width=2] (-0.2, 0.2) to (0.2, -0.2);
    
        \draw[black!80!white, line width=2] (-4.35, -2.0) to (-3.7, -2.0);
        \draw[black!80!white, line width=2] (-4.0, -1.7) to (-3.7, -2.0) to (-4.0, -2.3);
    
        \draw[black!80!white, line width=2] (-2.33, -2.0) circle (0.32);
        \fill[black!1!white] (-2.1, -1.95) rectangle (-1.8, -2.1);
        \fill[black!80!white] (-2.22, -1.95) to (-1.98, -1.95) to (-1.98, -1.7) to cycle;
    
        \draw[black!80!white, line width=2] (-0.88, -2.35) to (-0.88, -2.0) to (-0.95, -2.0) to (-0.65, -1.7) to (-0.35, -2.0) to (-0.43, -2.0) to (-0.43, -2.35) to cycle;
    
        \draw[black!80!white, line width=2.5] (1.65, -0.35) to (1.65, 0.35);
        \draw[black!80!white, line width=2.5] (1.3, 0.0) to (2.0, 0.0);
    
        \node[font={\Huge\bfseries\sffamily}, black!80!white] at (-3.37, 0.0) {New Tab};
    
        \node[font={\Huge\bfseries\sffamily}, black!50!white] at (2.5, -2.1) {Search};
    
        \draw[red, line width=3] (-0.4, -0.4) rectangle (0.4, 0.4);
        \draw[red, line width=3] (1.15, -0.5) rectangle (2.15, 0.5);
        \draw[red, line width=3] (-4.95, -0.8) rectangle (0.9, 0.9);
        \draw[red, line width=3] (-4.5, -2.5) rectangle (-3.5, -1.5);
        \draw[red, line width=3] (-2.8, -2.5) rectangle (-1.83, -1.5);
        \draw[red, line width=3] (-1.15, -2.5) rectangle (-0.15, -1.5);
        \draw[red, line width=3] (4.95, -1.15) rectangle (0.6, -2.95);
    
        \draw[black] (-5, 3) rectangle (5, -3.5);

    \end{scope}

    \draw[->, line width=4] (6, 0) to (7, 0);
    
    \begin{scope}[shift={(8.5, 0)}]
        \fill[black!1!white] (-0.4, -0.4) to (0.4, 0.4);
        \draw[black!80!white, line width=2] (-0.2, -0.2) to (0.2, 0.2);
        \draw[black!80!white, line width=2] (-0.2, 0.2) to (0.2, -0.2);
        \draw[red, line width=3] (-0.4, -0.4) rectangle (0.4, 0.4);
    \end{scope}


    
\end{tikzpicture}}
    \caption{The algorithm for extracting an image patch of the button which was clicked. First, the screen is captured in the proximity of the cursor. Subsequently, a \texttt{YOLOv5} model for button detection is used to segment all buttons on the image. The smallest button which includes the cursor is distinguished to be the one which was clicked.}
    \label{fig:extraction}
\end{figure}
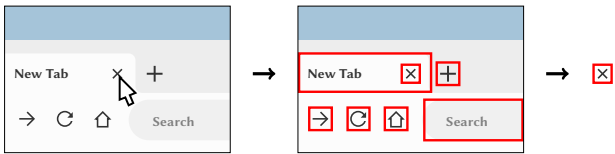

Furthermore, every click of an interactive area on the screen is identified as a user action too. To be able to reenact the click and also identify clicks on the same button with the same user action, we extract an image patch of the interactive area which we then store. The extraction of the image patch is done with the help of the \textit{YOLOv5} object detection model which was trained to detect buttons, links, and text fields on computer screens \cite{YOLOv5}. The extracted image patch is then matched against a database which stores an image patch for every interactive area a user has clicked on. The matching is performed by computing the normalized correlation coefficient \cite{OpenCVTemplateMatching}. If for some positioning of one image patch within another a high enough correlation exists, we have a match. It is necessary to first extend one of the image patches by a margin of a few pixels to account for inconsistencies in the extraction of image patches. To make the system DPI-independent -- particularly for multiple monitor setups -- we rescale the image patch to the default DPI of 96 beforehand. Additionally, to speed up this matching, some basic features like height, width, and mean color are computed for each image patch, which can be compared relatively quickly and reduce the amount of image-image correlations which need to be computed to determine if two image patches represent the same button. In case a match for the current image patch is found in the patch database, we need not add it again, but instead only associate it with the corresponding user action. The procedure of converting clicks to user actions is illustrated in \reffig{fig:extraction}.

To supply additional information connected to each observed user actions, we collect information about the system's state during it. The contextual information supplied to the network is: the application the action was taken in, in one-hot encoded representation; the relative $x$- and $y$-position of the cursor within the window the action was taken in; and the elapsed time since the last action was taken in multiples of 10 s, capped at 30 s to avoid this value from blowing up. 

\section{Model training}

We acquire data by recording a single user's activity while working on a Windows 10 dual monitor setup for a week. A brief description of the data can be found in \reftab{tab:training}.

\begin{table}[ht]
    \caption{Some quantities characterizing the data set.}
    \label{tab:training}
    \small
    \centering
    \begin{tabular}{ll}
        \toprule
        Quantity & Value \\
        \midrule
        Number of hours recorded & 46.21 \\
        Total number of user actions taken & 86'284 \\
        Number of distinct user actions encountered & 3'008 \\
        ~$\hookrightarrow$ which are clicks on interactive areas & 2'755 \\
        ~$\hookrightarrow$ which are keyboard, mouse, or combined input & 253 \\
        Storage space of data & 80.7 MB \\
        \bottomrule
    \end{tabular}
\end{table}

We only learn those user actions which we have encountered while acquiring the data set. Furthermore, interactive areas which were not clicked more than five times can be considered as rare and are ignored. Under these two simplifications we end up with an action space containing 442 distinct user actions.

\begin{figure}[ht]
    \centering
    \scalebox{0.3}{\begin{tikzpicture}

    \draw[->, line width=3] (-3, -1.5) to (-1, -3.5);
    \node[rotate=-45, font={\huge\bfseries\sffamily}] at (-2.25, -2.75) {time};
    
    \fill[lightblack, opacity=0.05] (-3, -1) rectangle (1, 5);
    \draw[white, line width=4] (-3, -1) rectangle (1, 5);
    \draw[white, line width=2] (-3, 1) rectangle (1, 2);

    \fill[lightblack, opacity=0.3] (-2.5, -1.5) rectangle (1.5, 4.5);
    \draw[white, line width=4] (-2.5, -1.5) rectangle (1.5, 4.5);
    \draw[white, line width=2] (-2.5, 1.5) rectangle (1.5, 1.5);
    
    \fill[lightblack, opacity=0.6] (-2, -2) rectangle (2, 4);
    \draw[white, line width=4] (-2, -2) rectangle (2, 4);
    \draw[white, line width=2] (-2, 1) rectangle (2, 1);

    \fill[lightblack, opacity=0.85] (-1.5, -2.5) rectangle (2.5, 3.5);
    \draw[white, line width=4] (-1.5, -2.5) rectangle (2.5, 3.5);
    \draw[white, line width=2] (-1.5, 0.5) rectangle (2.5, 0.5);

    \fill[lightblack] (-1, 0) rectangle (3, 3) node[midway, white, yshift=10pt, font={\Huge\bfseries\sffamily}] {user} node[midway, white, yshift=-10pt, font={\Huge\bfseries\sffamily}] {action};

    \begin{scope}[shift={(2.4, 0.9)}, scale=0.25]
        \draw[white, line width=2.5pt] (1, -0.5) to (1, 0.5) arc(0:180:1) to (-1, -0.5) arc(-180:0:1) to cycle;
        \draw[white, line width=3pt] (1, 0.4) to (-1, 0.1);
        \draw[white, line width=3pt] (0, 0.2) to (0, 1.5);
    \end{scope}

    \begin{scope}[shift={(-0.2, 1.8)}, scale=0.3]
        \draw[white, line width=2.5pt] (1.5, -0.5) to (1.5, 0.5) arc(0:90:0.5) to (-1, 1) arc(90:180:0.5) to (-1.5, -0.5) arc(180:270:0.5) to (1, -1) arc(270:360:0.5) to cycle;
        \draw[white, line width=3pt] (0.5, -0.5) to (-0.5, -0.5);
        \draw[white, line width=3pt] (-1.15, 0) to (-0.85, 0);
        \draw[white, line width=3pt] (-0.65, 0) to (-0.35, 0);
        \draw[white, line width=3pt] (-0.15, 0) to (0.15, 0);
        \draw[white, line width=3pt] (0.65, 0) to (0.35, 0);
        \draw[white, line width=3pt] (1.15, 0) to (0.85, 0);
        \draw[white, line width=3pt] (-1.15, 0.5) to (-0.85, 0.5);
        \draw[white, line width=3pt] (-0.65, 0.5) to (-0.35, 0.5);
        \draw[white, line width=3pt] (-0.15, 0.5) to (0.15, 0.5);
        \draw[white, line width=3pt] (0.65, 0.5) to (0.35, 0.5);
        \draw[white, line width=3pt] (1.15, 0.5) to (0.85, 0.5);
    \end{scope}

    \begin{scope}[shift={(2.1, 2.2)}, scale=1.0]
        \fill[white] (-0.3, -0.3) rectangle (0.3, 0.3);
        \draw[black!80!white, line width=2] (-0.2, -0.2) to (0.2, 0.2);
        \draw[black!80!white, line width=2] (-0.2, 0.2) to (0.2, -0.2);
    \end{scope}

    \fill[lightblack] (-1, 0) rectangle (3, -3) node[midway, white, yshift=10pt, font={\Huge\bfseries\sffamily}] {system} node[midway, white, yshift=-10pt, font={\Huge\bfseries\sffamily}] {state};
    \draw[white, line width=4] (-1, -3) rectangle (3, 3);
    \draw[white, line width=2] (-1, 0) to (3, 0);

    \node[white, font={\huge\bfseries\sffamily}] at (-0.15, -2.4) {$(x, y)$};

    \fill[white] (2.3, -1.8) circle (0.3);
    \fill[lightblack] (2.3, -1.8) circle (0.17);
    \draw[lightblack, line width=2pt] (2.3, -1.665) to (2.6, -1.665);
    \draw[lightblack, line width=2pt] (2.4, -1.9) to (2.25, -2.15);
    \draw[lightblack, line width=2pt] (2.22, -1.9) to (1.9, -1.55);
    \fill[white] (2.3, -1.8) circle (0.1);

    \fill[white] (-0.7, -0.3) rectangle (-0.1, -0.7);
    \fill[white] (-0.45, -0.7) rectangle (-0.35, -0.8);
    \fill[white] (-0.55, -0.8) rectangle (-0.25, -0.85);
    
    \draw[->, line width=3] (3.5, 0) to (4.5, 0);

    \fill[lightblack] (5, -1.5) rectangle (8, 1.5) node[midway, white, font={\Huge\bfseries\sffamily}] {RNN};

    \draw[->, line width=3] (8.5, 0) to (9.5, 0);
    
    \fill[lightblack] (10, -1.5) rectangle (13, 1.5) node[midway, white, font={\Huge\bfseries\sffamily}] {MLP};

    \draw[->, line width=3] (13.5, 0) to (14.5, 0);
    
    \fill[lightblack] (15, -1.5) rectangle (18, 1.5) node[midway, white, font={\Huge\bfseries\sffamily}] {Softmax};

    \draw[->, line width=3] (18.5, 0) to (19.5, 0);

    \fill[lightblack] (20, -1) rectangle (20.4, -0.5);
    \fill[lightblack] (20.5, -1) rectangle (20.9, -0.2);
    \fill[lightblack] (21, -1) rectangle (21.4, 0.5);
    \fill[lightblack] (21.5, -1) rectangle (21.9, 2.3);
    \fill[lightblack] (22, -1) rectangle (22.4, 1);
    \fill[lightblack] (22.5, -1) rectangle (22.9, 1.5);
    \fill[lightblack] (23, -1) rectangle (23.4, -0.1);
    \fill[lightblack] (23.5, -1) rectangle (23.9, -0.8);
    \fill[lightblack] (24, -1) rectangle (24.4, -0.9);
    \fill[lightblack] (24.5, -1) rectangle (24.9, -0.4);

    \node[font={\Huge\bfseries\sffamily}] at (22.5, -1.7) {likelihoods for next};
    \node[font={\Huge\bfseries\sffamily}] at (22.5, -2.6) {user actions};

\end{tikzpicture}}
    \caption{The architecture of the model used to predict the likelihood of every possible user action being taken.}
    \label{fig:architecture}
\end{figure}
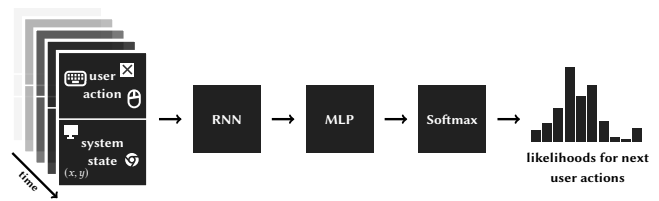

We train two RNNs, the LSTM and GRU on the time series of the one-hot encoded user actions coupled with the system information we mentioned above. At the end of the RNN, we append a shallow multilayer perceptron (MLP) with a Softmax output activation (see \reffig{fig:architecture}). Based on the last $n$ steps in the time series, the model is trained to predict a probability distribution over all user actions. This allows us to predict which user action most likely continues the sequence and additionally gives us information about the probability with which each user actions will be taken next. It also gives us an easy way of removing actions we are not interested in from the predictions: if we only care about predictions of interactive areas, we set the predicted probabilities corresponding to all other user actions to zero, renormalize the distribution, and then deduce the most likely actions from this filtered distribution. 

We fit the models using a sliding window along the time series of user actions: based on five preceding actions, we let the model predict the probability distribution over which action happens next and backpropagate the error measured with the cross-entropy loss. The parameters of the models are summarized in \reftab{tab:RNN}.

\begin{table}[ht]
    \caption{Explanation of the parameters of the RNN.}
    \label{tab:RNN}
    \small
    \centering
    \begin{tabular}{lll}
        \toprule
        Parameter & Value & Explanation \\
        \midrule
        \texttt{n\_past} & 5 & Number of past user actions \\ 
        \texttt{hidden\_size} & 600 & Number of hidden features \\ 
        \texttt{num\_layers} & 1 & Number of stacked RNN cells \\
        \texttt{loss\_function} & \texttt{CrossEntropyLoss} & Loss function that is minimized \\
        \texttt{optimizer} & \texttt{Adam} & Optimizer used for training \\ 
        \texttt{learning\_rate} & 0.001 & Learning rate of optimizer \\
        \bottomrule
    \end{tabular}
\end{table}

\section{Results}
\label{sec:results}

\begin{figure}[ht]
    \centering
    \scalebox{0.6}{\input{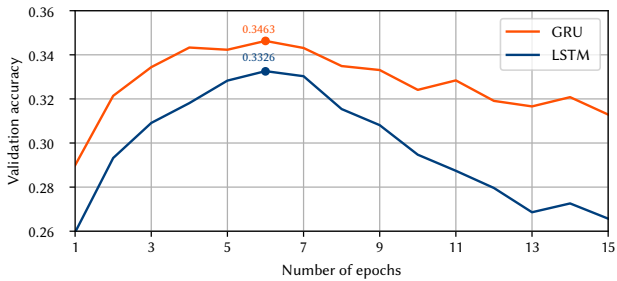}}
    \caption{The per-epoch accuracy of the two RNN architectures for predicting the next user action on an unseen validation set of 6'000 actions (approximately one day of user activity). The maximum accuracy of 34.63 \% of the GRU is attained after just 256.8 s of training on a single 12\textsuperscript{th} generation Intel i7 processor.}
    \label{fig:accuracy}
\end{figure}

It turns out that even rather small RNNs are able to deliver reasonably good predictions. The validation accuracy, i.e. the relative amount of correct next action predictions on the validation set, is plotted in \reffig{fig:accuracy}. In this context, a prediction is said to be correct if the user action of highest predicted probability corresponds exactly to the action the user ended up taking. In case the model encounters a user action which it has not seen during training, it ignores it while making a prediction.

\begin{figure}[ht]
    \centering
    \scalebox{0.2}{\input{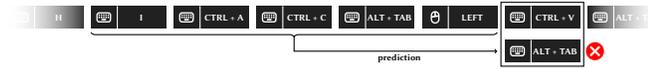}}
    \caption{Visualization of how the accuracy of a predictor is determined on the validation set. From the series of user actions, five consecutive user actions are sampled, and the most likely next action is predicted and compared to the action the user ended up taking. In this example, the user action of highest probability (ALT $+$ TAB) based on the previous five user actions did not correspond to what the user did (CTRL $+$ V).}
    \label{fig:validation}
\end{figure}

\begin{figure}[ht] 
    \centering
    \scalebox{0.63}{\input{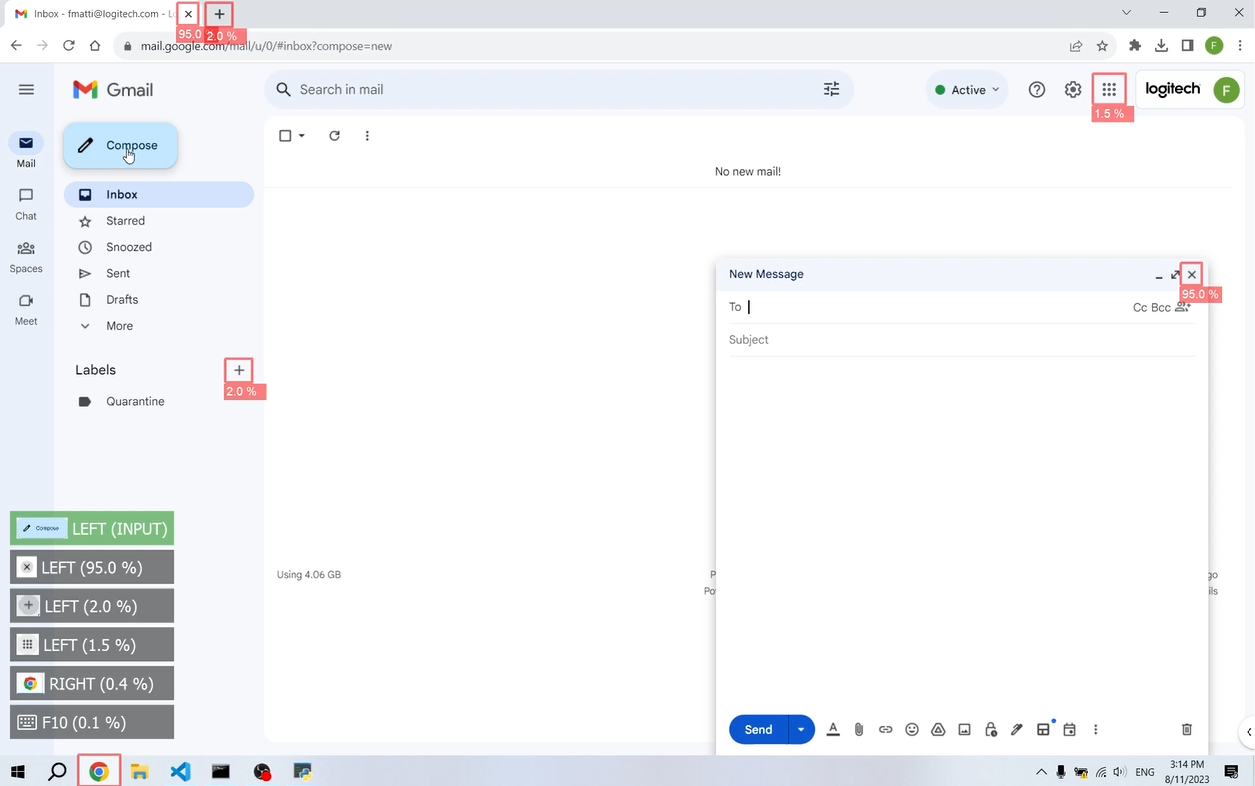}}
    \caption{Screenshot of the application which predicts and visualizes the five next most likely actions while a user is working on the computer.}
    \label{fig:screenshot}
\end{figure}

To demonstrate the capability of predicting actions in real-time, we built a small application which listens to a user's inputs, predicts what the user is most likely going to do next based on their previous usage of the computer, and finally visualizes the five most likely actions on the screen. This includes finding the predicted buttons on the screen and highlighting them with a red box; by far the most time-consuming task of the system. To locate an image patch on the screen we compute the normalized correlation coefficient for every possible arrangement of the image patch on the screen \cite{OpenCVTemplateMatching}. In the location where the correlation exceeds a certain threshold, we have a match. A screenshot of this program in use can be found in \reffig{fig:screenshot}.

\section{Discussion}
\label{sec:discussion}

The above work showcases a system which predicts how likely each action is going to be taken by a computer user. To do so, we record a user's keyboard and mouse inputs which we subsequently convert to a novel representation, termed user actions. We then train recurrent neural networks (RNN) to predict a probability distribution over how likely every possible action follows a sequence of previously taken actions. We show that after less than five minutes of training a small gated recurrent unit (GRU), 34.63 \% of the predicted user actions coincide with the one which is indeed taken by the user. To put this into perspective, out of almost 500 considered user actions, one week of user data and minimal training is enough to predict the user's next action in more than a third of the cases. It turns out that the model is even able to perform very basic text completion tasks. However, there exist way more suitable representations for these problems \cite{LLM}, which we decided not to exploit due to their little value for the applications we have in mind.

\begin{figure}[ht]
    \centering
    \scalebox{0.35}{\begin{tikzpicture}
    \draw[->, line width=5pt, black!20!white] (15.15, 2.4) to (13, 4.5);
    \draw[->, line width=5pt, black, out=125, in=20] (15.15, 2.4) to (12.5, 3);

    \draw[rounded corners=35pt, line width=2pt, opacity=0.3] (0, -0.25) rectangle (10, 4.25);
    \draw[rounded corners=45pt, line width=2pt, opacity=0.2] (-0.5, -0.75) rectangle (10.5, 4.75);
    \draw[rounded corners=55pt, line width=2pt, opacity=0.1] (-1, -1.25) rectangle (11, 5.25);
    
    \fill[rounded corners=25pt, lightblue] (0.5, 0.25) rectangle (9.5, 3.75) node[midway, xshift=7ex, yshift=-1ex, scale=1.7, font={\Huge\bfseries\sffamily}, black] {Compose};
    \draw[black, line width=3pt] (2.4, 2.4) to (1.6, 1.6) to (1.6, 1.4) to (1.8, 1.4) to (2.6, 2.2);
    \draw[black, fill=black, line width=3pt] (2.4, 2.4) to (2.7, 2.7) to (2.9, 2.5) to (2.6, 2.2);
    
    \begin{scope}[shift={(0.8,4.1)}]
        \draw[black!5!white, fill=white, line width=2.5pt] (15, -2) to (15, -3) to (15.25, -2.9) to (15.4, -3.25) to (15.6, -3.15) to (15.45, -2.8) to (15.7, -2.7) to cycle;
    \end{scope} 
    \begin{scope}[shift={(0.5,4.3)}]
        \draw[black!20!white, fill=white, line width=2.5pt] (15, -2) to (15, -3) to (15.25, -2.9) to (15.4, -3.25) to (15.6, -3.15) to (15.45, -2.8) to (15.7, -2.7) to cycle;
    \end{scope} 
    \begin{scope}[shift={(0.2,4.5)}]
        \draw[black!50!white, fill=white, line width=2.5pt] (15, -2) to (15, -3) to (15.25, -2.9) to (15.4, -3.25) to (15.6, -3.15) to (15.45, -2.8) to (15.7, -2.7) to cycle;
    \end{scope}
    \begin{scope}[shift={(-0.1,4.7)}]
        \draw[black, fill=white, line width=2.5pt] (15, -2) to (15, -3) to (15.25, -2.9) to (15.4, -3.25) to (15.6, -3.15) to (15.45, -2.8) to (15.7, -2.7) to cycle;
    \end{scope}
    

\end{tikzpicture}}
    \caption{Sketch of how the mouse cursor is attracted to a button.}
    \label{fig:attraction}
\end{figure}
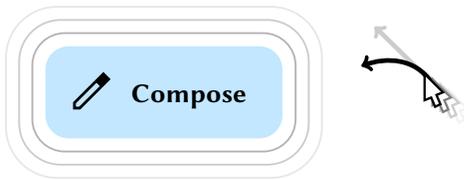

We can think of many ways in which the predictions for the next action can be leveraged to improve computer users' speed or precision: automatically completing repetitive input based on a user's previous activity; making the most frequently used buttons accessible for visually impaired people; and attracting the cursor to the buttons the user is most likely going to click, to name just a few. To achieve the last, we manipulate the position of the cursor on the screen in such a way that it is attracted by, i.e. pulled towards, the buttons on the screen which the user is most likely going to click (see \reffig{fig:attraction}). To do so, every time the user takes an action we use a trained RNN to predict the buttons which the user is most likely going to click next. We then load the image patch from the database of all clicked buttons and locate it on the screen. The attraction is implemented by treating the cursor as an object moving in a gravitational field induced by the buttons. Instead of mass, the confidence with which the button is going to be clicked determines the gravitational pull. Now, every time the cursor is moved, we modify its position based on the gravitational pull it feels from the buttons.

A small limitation of our system is produced by buttons which change size or shape. For instance, there are some applications where buttons are wider when working in full screen mode than in windowed mode. It is clear that buttons of different shapes will not yield a match in our image patch comparison routine, and will be considered as two different buttons. Other buttons significantly change their appearance when hovering the cursor over them, such that they will fail to yield a desired match when comparing it to other image patches. Problematic are also buttons with notification badges displayed on top of them.

Nevertheless, we hope the current work has managed to demonstrate the feasibility, and the value, of predicting mouse and keyboard actions in real-time. We also hope this study will be further expanded and serve as a reference for future research in this domain.

\bibliographystyle{ACM-Reference-Format}
\bibliography{sample-base}

\end{document}